\documentclass[12pt]{article}
\pdfoutput=1
\usepackage{subfigure}
\usepackage{graphicx}
\usepackage{amssymb,amsmath,mathrsfs,bm}
\usepackage{hyperref}
\numberwithin{equation}{section}
\begin{document}
\begin{titlepage}
\vspace{-3.0cm}
\begin{flushright}
CERN-PH-TH/2013-149
\end{flushright}
\begin{center}
\begin{Large}
{\bf Modular discretization

\hskip2.0truecm

of the AdS$_2$/CFT$_1$ Holography}
\end{Large}

\vskip1.0truecm

{\large{Minos Axenides$^{1}$, Emmanuel Floratos$^{1,2,3}$ and Stam Nicolis$^{4}$}}

\vskip0.5truecm

 {\sl $^1$ Institute of Nuclear and Particle Physics, N.C.S.R. Demokritos, GR-15310, Athens, Greece}

\vskip0.15truecm

{\sl $^2$  Department of Physics, Univ. of Athens, GR-15771 Athens, Greece}

\vskip0.15truecm

{\sl $^3$ Department of Physics, CERN Theory Division,  CH-1211 Geneva 23, Switzerland}

\vskip0.15truecm

{\sl $^4$ CNRS-Laboratoire de Math\'{e}matiques  et  Physique  Th\'{e}orique (UMR 7350)\\
F\'{e}d\'{e}ration \ `Denis \ Poisson' (FR2964)\\
Universit\'{e} \ de \ Tours \ `Fran\c{c}ois  Rabelais', Parc \ Grandmont, \ 37200  \ Tours, France}\\

\vskip0.1truecm

{axenides@inp.demokritos.gr; \ \ mflorato@phys.uoa.gr; \ \ Stam.Nicolis@lmpt.univ-tours.fr}
\end{center}

\vspace{0.1truecm}

\begin{abstract}


{\small
\noindent
  We propose a finite discretization  for the black hole geometry and dynamics. We realize our proposal, in the case of extremal black holes, for which the radial and temporal near horizon geometry is known to be  AdS$_2 =SL(2,\mathbb{R})/SO(1,1,\mathbb{R})$.  We implement its discretization  by replacing  the set  of real numbers $\mathbb{R}$  with the set of integers modulo  $N$
with AdS$_2$ going over to the finite geometry  AdS$_2[N] =SL(2,\mathbb{Z}_N)/SO(1,1,\mathbb{Z}_N)$.
 We model the dynamics  of the  microscopic degrees of freedom  by  generalized  Arnol'd cat maps ${\sf A} \in SL(2,\mathbb{Z}_N)$,  which are isometries of the geometry at both the classical and quantum levels.
  These exhibit well studied properties  of strong arithmetic chaos, dynamical entropy, nonlocality and factorization  in the cutoff discretization $N$, which are crucial for fast quantum information processing.
 We construct, finally,  a new kind of  unitary and holographic correspondence, for AdS$_2[N]$/CFT$_1[N]$,  
via  coherent states of both the  bulk and boundary geometries.}

 \end{abstract}

\end{titlepage}

\tableofcontents
\section{Introduction}\label{intro}
From the early days of the black hole information paradox  it has been suggested  that,  in order to  bring together quantum mechanics and the equivalence principle consistently, we have to give up locality and the semiclassical description in the near horizon region. Key ingredients in the framework,  that is likely to replace them,  include  the principle of holography~\cite{thooftBHQM,preskillbhInfo}, complementarity~\cite{susskindbhComp} and strongly chaotic dynamics~\cite{susskindscramble}.
Recent discussions on  the nature and the dynamics of microscopic degrees of freedom in the near  horizon region of black holes have, also, strengthened the conviction that semiclassical physics is inadequate to guarantee the compatibility of the  equivalence principle  and quantum mechanics.Therefore it seems imperative that a drastic departure from conventional, semiclassical, physics is needed ~\cite{AMPSS}.

We should stress here that recent developments in string theory~\cite{Sen12,Sen},  which take into account the correct definition of the black hole entropy at the quantum level, have improved considerably our understanding of the black hole microscopic degrees of freedom. We understand now some important quantum statistical  properties,  in particular,  the exact black hole quantum entropy,  for a certain class of extremal black holes. 

Important issues have  remained open,  which dominate the recent literature. They pertain to  the description of the nature and dynamics of the near horizon microstates. 

In what follows we consider  the simplest dynamical context for the discussion of the black hole information paradox \cite{Susskind-Lindesay}.
 Consider two observers, one at infinity, $O_a$ and the other in free fall, $O_b$ into the black hole. The question is, if there exists a unitary transformation,  connecting the description of a freely falling particle near the horizon by the two observers.

In the seminal paper~\cite{dAFF} the corresponding quantum mechanical evolution operators have been constructed, by two, different, conformal invariant, Hamiltonians. 
 The ``first'' Hamiltonian, which, in our case, describes the time evolution for observer $O_a$, has continuous spectrum, while the ``second'' Hamiltonian, which describes the time evolution for observer $O_b$, in our case, has discrete spectrum.

 There has been extensive discussion in the literature,  how these two descriptions can be related to the paradox between the infinite number of states near the horizon thus described and the finite entropy of the black hole, which is proportional to its area~\cite{Kallosh,Strominger,townsendetal}. Though developments in string theory~(reviewed in \cite{Sen12}), around the same time, succeeded in resolving the problem of counting these states and of reproducing the thermodynamical result for the black hole entropy,  how to extend this calculation for the case of a ``small'' number of microstates, i.e. for ``small'' black holes, remains open~\cite{Sen12}. 
 
   The information paradox, in the case of extremal black holes, which have an AdS$_2$ radial and temporal, near horizon, geometry,  is expressed by the putative mismatch between the description accessible to observer, $O_a$, on the boundary, using a CFT$_1$ dynamics and that of the, free--falling, observer, $O_b$, who is using the bulk, AdS$_2$, dynamics. The resolution of this paradox is currently the subject of intense research activity~\cite{paradoxADSCFT}.
 
 In the present  and subsequent works we  study the above issues using  a space-time discretization of the near horizon geometry, $\mathrm{AdS}_2 =SL(2,\mathbb{R})/SO(1,1,\mathbb{R})$, of extremal black holes.
We assume it to be discrete and finite as a consequence of the finite dimension of the Hilbert space of black hole microstates.  Indeed, the existence of a finite number of linearly independent wavefunctions as  probes implies  a finite resolution of its spacetime geometry.
This is achieved  by replacing  the set  of real numbers $\mathbb{R}$  by the set of integers modulo $N$, $\mathbb{Z}_N$, for any positive integer $N$. The  discretization thus replaces the continuous spacetime  AdS$_2$ by the finite arithmetic geometry,  AdS$_2[N] =SL(2,\mathbb{Z}_N)/SO(1,1,\mathbb{Z}_N)$~\cite{terras,finitegeometry}.

This discretization, which we call ``modular discretization'', has the merit of preserving corresponding symmetry properties of the continuous, classical, space-time geometry and provides a means of defining a consistent quantum geometry, through holography. In this discretized setting we will construct the corresponding unitary evolution operators for the two observers.

The discretization defines an infrared cutoff $L$ as well as a UV one $ \frac{L}{N}$. 
We obtain a discretized spacetime by considering the lifting to the AdS$_{2}$  of the $L\times L$ square ligth cone lattice by stereographic projection. It is obvious that
the continuum limit can be recovered by taking first the $N\to \infty$ limit at fixed $L$ and, afterwards, the limit $L \to \infty$. 
It is important to stress at this point the independence of the cutoff, $L$
from  the AdS$_2$ radius $R_{\mathrm{AdS}_2}$.  

In order to describe  the dynamics of probes, at both the classical and quantum level,  we use the  Arnol'd cat maps ${\sf A}$, which are elements of $SL(2,\mathbb{Z}_N)$.  
They are known to possess properties of strong arithmetic chaos, ergodic mixing and non-locality~\cite{Arnold,Vivaldi}. 
These maps also satisfy factorization properties in the discretization cutoff  $N$ , which  induce fast quantum information processing between the probe and the near horizon geometry~\cite{fastqmaps,entangledfqm}.

Our present  work builds on our earlier work on Finite Quantum Mechanics (FQM)~\cite{floratos89}. Therein we introduced  the discretized toroidal membrane  $\mathbb{Z}_N\times \mathbb{Z}_N$  as a tool for studying the
Matrix model truncation of the membrane dynamics~\cite{BFSS}. It renders the discrete membrane as a quantum phase space of finite quantum mechanics, which possesses the canonical transformation group $SL(2,\mathbb{Z}_N)$~\cite{Berry,Ford,balian_itzykson,athanasiu_floratos,afnholo}. 

Interestingly enough, the discretized membrane, in the black hole setting, describes the geometry of the stretched horizon~\cite{Susskind-Lindesay} and the Matrix model describes the M--theoretic dynamics of its microscopic degrees of freedom~\cite{susskindscramble}.

We  extend these results to the case of  the AdS$_2[N],$ 
discrete, near horizon, bulk geometry, and the dynamics of the infalling observer, $O_b$, 
along with its associated boundary CFT$_1[N]$ and the observer $O_a$ in order to obtain a holographic correspondence. In the discrete case the boundary is  constructed as a coset space $SL(2,\mathbb{Z}_N)/\mathfrak{B}_N$, which is identified with the discrete projective line, $\mathbb{RP}_N^1$. 
Here $\mathfrak{B}_N$ is the Borel subgroup of $SL(2,\mathbb{Z}_N)$, which fixes the point at infinity. 

The group $SL(2,\mathbb{Z}_N)$, in the present context,   plays three different roles: 
a)as the isometry group of AdS$_2[N]$,  b)
as the symplectic group of AdS$_2[N]$, which is considered to be a (stringy) phase space and 
c) as  the conformal group of the boundary.  Properties (a) and ( c ) are the  basic reasons for the existence of the AdS$_2[N]$/CFT$_1[N]$ correspondence and (b) will be used for the quantization of both the geometry(states) and the dynamics(evolution operator). 

We construct the discrete time evolution, quantum unitary maps  explicitly  and discuss their action  
on the common $N-$dimensional Hilbert space of both the bulk and the boundary.
The natural action of these quantum maps is realized on the set of coherent states, appropriate for the bulk and boundary coset geometries.
These states inherit classical chaotic dynamics, define isometric invariant (bulk-bulk, bulk-boundary and boundary-boundary) propagators and  are convenient for the discussion of  localization problems 
in the  AdS/CFT correspondence, since they saturate the  uncertainty relation
of the UV/IR connection ~\cite{adscft}.

The plan of the present work is as follows:

 In section~\ref{LCWeyl} we review
the construction of the smooth AdS$_2$ geometry and its boundary, as a doubly ruled surface by rotating the light
cone lines around the time-like circle. We establish  various global coordinate
systems, through appropriate coset parametrizations.
  More specifically we show that the light cone coordinates, on the stereographic projection plane, 
	parametrize holographically both the bulk and its boundary.

In order to describe the high energy dynamics for the the radial motion of probes,   we employ  linear isometry maps,  ${\sf A}\in SL(2,\mathbb{R})$ , which are appropriate  for the description of the infalling (bulk) and static (boundary) observers.

In section~\ref{HorModN} we motivate the introduction of the arithmetic discretization $\mathrm{mod}\,N$. We define the Finite Quantum Mechanics for both the bulk and the boundary on the same Hilbert space.
We shall work in the Hilbert space of the metaplectic representation of $SL(2,\mathbb{Z}_N)$ of dimension $N$ for the simplest case $N=p$ of an odd prime. In this case $Z_{N}=\mathbb{F}_p$ is the  simplest Galois  field.

The methods to be presented apply also for all other irreps of this group. 
In the case $N=p$, an odd prime, the number of irreps is $p+4$. They have been worked out in detail , using the method of induced representations, which correspond to the multiplicative characters of $\mathbb{F}_p$ ,or $\mathbb{F}_{p^2}$~\cite{Silberger}.

The boundary is also constructed as a coset space $SL(2,\mathbb{F}_p)/\mathfrak{B}_p$, which is identified with the discrete projective line, $\mathbb{RP}_p^1$. Here $\mathfrak{B}_p$ is the Borel subgroup of $SL(2,\mathbb{F}_p)$, which fixes the point at infinity.

In section~\ref{CohStates} we explicitly construct the bulk and the boundary overcomplete set of discrete coherent states. We discuss their basic properties as they are appropriate for the corresponding coset geometries.

The states and the observables are also expanded on the coherent states and their time evolution is defined through the quantum cat maps.
The correlation functions of various observables are defined, as well as the method of their evaluation.

Finally, in section~\ref{Hologcat},  we exhibit the reconstruction (holography) of the bulk coherent states from those of the boundary, via the
bulk-boundary, bulk-bulk, boundary-boundary propagators and the
consequent reconstruction of the scalar  bulk observables from the boundary ones. The correlation functions of scalar observables in the bulk and the boundary are connected through this holography which can be explicitly calculated.

In the last section~\ref{Concl} we summarize  our results and their role in the context of the problem of black hole information  processing. We also comment on future work on the complete description of finite conformal quantum mechanics on the boundary as well as on how the scrambling time bound might be saturated~\cite{susskindscramble}.

\section{Observers, geometry of cosets and Weyl dynamics  on AdS$_2$ }\label{LCWeyl}

Consider the  dynamics of freely falling bodies, in the near horizon region of  spherically symmetric 4d extremal black holes. The geometry  is known to be of the form
AdS$_2\times S^2$, where the AdS$_2 =SL(2,\mathbb{R})/SO(1,1,\mathbb{R})$,  factor describes the geometry of the radial and time coordinates
 and  $S^2$  is  the    horizon surface.
We will compare the description of high energy radial dynamics as seen by (radial) observers(static or freely falling), for which the transverse and longitutinal motion is decoupled.

To each of these observers corresponds  a global  space-time coordinate system and in the following we shall exhibit some of them using
group theory.

The AdS$_2$ spacetime,  is a one-sheeted hyperboloid defined through its
global embedding in Minkowski spacetime with one space-- and two time--like
dimensions, ${\mathscr M}^{1,2}$,  by the equation~\cite{Gibbons,Bengtsson}.

\begin{equation}
\label{AdS2_M21}
x_0^2 + x_1^2 - x_2^2 = 1
\end{equation}

The boundaries of AdS$_2$ consist of two time--like disconnected circles, where
AdS$_2$ approaches  asymptotically the light cone of ${\mathscr M}^{1,2}$
\begin{equation}
\label{M21_LC}
x_0^2 + x_1^2 - x_2^2 = 0
\end{equation}

AdS$_2$ is at the same time the homogeneous space
 $SO(1,2)/SO(1,1)$.  This case
  is special in that $SO(1,2)$ has a double cover, $SL(2,\mathbb{R}),$ so we  have $\mathrm{AdS}_2 =
SL(2,\mathbb{R})/SO(1,1)$.

In order to establish our notation and conventions, we proceed with the Weyl construction of the
double covering group,  $SL(2,\mathbb{R})$.

To every point $x_\mu\in\mathrm{AdS}_2$, $\mu=0,1,2$,  we assign the
traceless and real,  $2\times 2$ matrix

\begin{equation}
\label{Weyl}
{\sf M}(x)\equiv\left(\begin{array}{cc} x_0 & x_1+x_2 \\x_1-x_2 & -x_0\end{array}\right)
\end{equation}
Its determinant is
 $\mathrm{det}\,{\sf M}(x)=-x_0^2-x_1^2+x_2^2=-1$.

The action of any ${\sf A}\in SL(2,\mathbb{R})$ on AdS$_2$ is defined through the non-linear
mapping

\begin{equation}
\label{Weyl_mapping}
{\sf M}(x') = {\sf A}{\sf M}(x){\sf A}^{-1}
\end{equation}

 This induces an $SO(1,2)$
transformation on $(x_\mu)_{\mu=0,1,2}$,

\begin{equation}
\label{induced_transf}
x' \equiv {\sf L}({\sf A}) x
\end{equation}

Choosing as the origin of coordinates  the base point  $\bm{p}\equiv (1,0,0)$,  its
stability group  $SO(1,1)$ is the group of Lorentz transformations in the
$x_0=0$ plane of ${\mathscr M}^{1,2}$  or equivalently,  the ``scaling''
subgroup ${\sf D}$  of $SL(2,\mathbb{R})$

\begin{equation}
\label{scaling}
{\sf D}\ni {\sf S}(\lambda)\equiv \left(\begin{array}{cc} \lambda & 0 \\ 0 & \lambda^{-1}\end{array}\right)
\end{equation}
for $\lambda\in\mathbb{R}^\ast$.

For this choice of the stability point,  we define the coset $h_{\sf A}$ by decomposing  ${\sf A}$  as
\begin{equation}
\label{Acoset}
{\sf A} = h_{\sf A}{\sf S}(\lambda_{\sf A})
\end{equation}
Thus, we  associate uniquely to every point $x\in\mathrm{AdS}_2$
the corresponding coset representative $h_{\sf A}(x)$.

We introduce now  the global coordinate system  defined  by the
 straight lines that generate AdS$_2$  and for which it can be checked easily
that they form its complete  set of light cones.

\begin{figure}[thp]
\begin{center}
\includegraphics[scale=0.8]{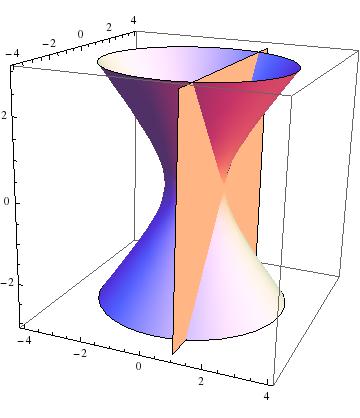}
\end{center}
\caption[]{The light cone of AdS$_2$ at $\bm{p}=(1,0,0)$.}
\label{LCAdS2}
\end{figure}

Consider the two lines,
 $\bm{l}_\pm(\bm{p})$,  passing through the point $\bm{p}\in{\mathscr M}^{1,2}$
orthogonal to the $x_0$ axis and at angles $\pm\pi/4$ to the $x_1=0$ plane. They are defined by the intersection of AdS$_2$ and the plane $x_0=1$~cf.~fig.~\ref{LCAdS2}.

The coordinates of any point, $\bm{q}_+\in\bm{l}_+(\bm{p})$ and
 $\bm{q}_-\in\bm{l}_-(\bm{p})$ are given  by  $(1,\mu_\pm,\pm\mu_\pm)$,
 $\mu_\pm\in\mathbb{R}$ respectively.

Rotating these lines  ,
 around the $x_0,x_1$ time circle  by appropriate angles
$\phi_\pm\in[0,2\pi)$ , we can parametrize any point by  their intersection with coordinates

\begin{equation}
\label{new_points}
\begin{array}{l}
\displaystyle
x_0 = \cos\phi_\pm -\mu_\pm\sin\phi_\pm\\
\displaystyle
x_1 = \sin\phi_\pm +\mu_\pm\cos\phi_\pm\\
\displaystyle
x_2 = \pm\mu_\pm
\end{array}
\end{equation}
  The corresponding pair of crossing lines ,$\bm{l}_\pm(\bm{x})$, define the local light cone.

Another form of the previous equation is:
\begin{equation}
\label{inverse_mapping}
\begin{array}{ccc}
\displaystyle
e^{\mathrm{i}\phi_\pm} = \frac{x_0\pm\mathrm{i}x_1}{1\pm x_2} &
\displaystyle &
\displaystyle
\mu_\pm = \pm x_2
\end{array}
\end{equation}

The corresponding coset parametrization (group coset motion which brings the origin to the point $x$) is:
\begin{equation}
\label{cosets}
h(\mu_\pm,\phi_\pm) = {\sf R}(\phi_\pm){\sf T}_\pm(\mu_\pm)
\end{equation}
where
\begin{equation}
\label{coset_rot}
{\sf R}(\phi) = \left(\begin{array}{cc} \cos\phi/2&
  -\sin\phi/2\\\sin\phi/2& \cos\phi/2\end{array}\right)
\end{equation}
and
\begin{equation}
\label{coset_trans}
{\sf T}_+(\mu) = \left[{\sf T}_-(-\mu)\right]^\mathrm{T} =
\left(\begin{array}{cc} 1 & -\mu\\ 0 & 1\end{array}\right)
\end{equation}
We notice  that ${\sf T}_\pm(\mu_\pm)$, acting on the base point
$X(\bm{p})$, generate the light cone~$l_\pm(\bm{p})$. Hence we identify these one parameter subgroups with the light cones at $p$.

At this point we should like to pause and discuss the physical interpretation of the rotation~(\ref{coset_rot}) and translation~(\ref{coset_trans}) groups.

 Consider the two observers, in the AdS$_2$ background.  One of them, $O_a$, is located at infinity  and the other one, $O_b$, is in free fall. Their corresponding classical description, for a freely falling particle near the horizon, is given, for $O_a$, by the finite group of translations, with time parameter $\mu$. For  observer $O_b$ it is given by the finite group of rotations, with time parameter $\phi$ .  In the seminal paper~\cite{dAFF} the corresponding quantum mechanical evolution operators have been constructed in order to describe models for confinement and asymptotic freedom, by two different conformal  Hamiltonians.
 The ``first'' Hamiltonian, which in our case describes the time evolution for observer $O_a$, has continuous spectrum. The ``second'' one which describes the time evolution for observer $O_b$,  has discrete spectrum.

After this intermezzo, we proceed with the description of AdS$_2$ as a phase space.

We observe that the  variables, $\phi$ and $\mu$, are in fact  Darboux coordinates
 with respect to the natural $SO(2,1)$ invariant Poisson structure, which promotes AdS$_2$ to a phase space. They are  conjugate variables
 and they parametrize the time evolution, at the quantum mechanical level, of the two observers, $O_a$ and $O_b$, thereby realizing the complementarity of the physics they describe~\cite{susskindbhComp}.

It is also possible to use the  light cone coordinates , $\mu_\pm$ in order to parametrize AdS$_2$, thereby
eliminating the angles $\phi_\pm$. The corresponding cosets are:
\begin{equation}
\label{translcoset}
h(\mu_+,\mu_-) = {\sf T}_-(\mu_-){\sf T}_+(\mu_+)
\end{equation}
which define a global light cone coordinate system. The map between
$(\mu_+,\mu_-)$ and $(x_0,x_1,x_2)$ is easily obtained:
\begin{equation}
\label{trans_map}
\begin{array}{ccc}
\displaystyle
\mu_+ = \frac{x_1 + x_2}{2} & \displaystyle\mathrm{and} &
\displaystyle
\mu_- = \frac{x_1 - x_2}{1+ x_0}
\end{array}
\end{equation}
The light cone cosets establish the causal patches of any observer on AdS$_2$ and thus
the causal diamonds of any pair of observers~\cite{Gibbons_Patricot}.

For completeness, we exhibit also the standard system of hyperbolic global coordinates,
\begin{equation}
\label{hyperbolic_global}
\begin{array}{ccccc}
\displaystyle x_0 = \cosh\psi\cos\chi, & \displaystyle &
\displaystyle x_1 = \cosh\psi\sin\chi, & \displaystyle &
\displaystyle x_2 = \sinh\psi
\end{array}
\end{equation}
and the corresponding   coset parametrization,

\begin{equation}
\label{coset_hyp}
h(\psi,\chi)={\sf R}(\chi){\sf H}(\psi)
\end{equation}
with
\begin{equation}
\label{hypcoset}
{\sf H}(\psi) = \left(\begin{array}{cc} \cosh\psi/2& \sinh\psi/2\\
\sinh\psi/2 & \cosh\psi/2\end{array}\right)
\end{equation}
an element of the Lorentz group that acts in the $x_1=0$ plane.

These coset parametrizations induce also  specific metrics on AdS$_2$. For the
parametrization~(\ref{cosets}) we obtain
\begin{equation}
\label{AdS2metric_phi_mu}
ds^2 = (1+\mu^2)d\phi^2 + 2d\phi d\mu
\end{equation}
Substituting in this expression $\mu\equiv\tan\sigma$,
$\sigma\in\left(-\pi/2,\pi/2\right)$, we obtain the Einstein strip
\begin{equation}
\label{Einstein_cylinder}
ds^2 = \frac{1}{\cos^2\sigma}\left[-d\tau^2 + d\sigma^2\right]
\end{equation}
with $\tau\equiv\sigma + \phi\in\mathbb{R}$ with the two disconnected boundaries at $\sigma\equiv\pm\pi/2$

For the standard hyperbolic global coordinate system,  $(\psi,\chi )$,  we obtain
the metric
\begin{equation}
\label{hyperbolic_metric}
ds^2 = -\cosh^2\psi\ d\chi^2 + d\psi^2
\end{equation}
with $\psi\in(-\infty,\infty)$ and $\chi\in[0,2\pi)$.

Finally, for the light cone coordinates $\mu\equiv\mu_+, \nu\equiv\mu_-$ of
eq.~(\ref{trans_map}) we find the metric
\begin{equation}
\label{trans_metric}
ds^2 = -4\left(\nu^2  d\mu^2 + \mu^2 d\nu^2 -d\mu d\nu\right)
\end{equation}


In the rest of this section we discuss the dynamics of probes,  appropriate for the description of the string ground state geometry.

The AdS$_2$ coset geometry, inherits  a  symplectic structure and
a non-degenerate Poisson bracket from the isometry group, given by
\begin{equation}
\label{poisson_bracket}
\begin{array}{ccccc}
\displaystyle \left\{x_0,x_1\right\} = -x_2 & \displaystyle &
\displaystyle \left\{x_1,x_2\right\} = x_0 & \displaystyle &
\left\{x_2,x_0\right\} = x_1
\end{array}
\end{equation}
These relations are realized, for example, in the global coordinate system
$(\phi,\mu)$, where the area element is $d\phi d\mu$ and the coordinates
$\phi$ and $\mu$ are  Darboux coordinates, as
\begin{equation}
\left\{f,g\right\} =
\frac{\partial f}{\partial\phi}\frac{\partial g}{\partial\mu} -
\frac{\partial f}{\partial\mu}\frac{\partial g}{\partial\phi}
\end{equation}
The corresponding Hamilton's equations for incompressible flows on $AdS_2$ are,
\begin{equation}
\label{poissonflow}
\dot{x}_\mu = \left\{x_\mu, H\right\}
\end{equation}

 The simplest classical  motions of probes of lowest energy  are described by the isometric maps ${\sf A}\in
SL(2,\mathbb{R})$.

At the level of discrete time evolution (maps), this isometric motion is parametrized as follows:
If, for instance,  $h(\phi,\mu)\in SL(2,\mathbb{R})$ is a coset, describing the probe's position in AdS$_2$,
at proper time $\tau$, then at time $\tau + 1$  it will evolve as
\begin{equation}
\label{one_step_op}
h(\phi_{\tau+1},\mu_{\tau+1})= {\sf A}h(\phi_\tau,\mu_\tau)(\mathrm{mod}\,{\sf D})
\end{equation}
Using the decompositions~(\ref{Acoset},\ref{cosets}),
the parameters $\phi_{\sf A},\mu_{\sf A},\lambda_{\sf A}$ in can be given explicitly,  in terms of the matrix elements of
$$
{\sf A} = \left(\begin{array}{cc} a & b\\ c & d\end{array}\right)
$$
by the expressions
\begin{equation}
\label{coset_params}
\begin{array}{ccccc}
\displaystyle
\cos\frac{\phi_{\sf A}}{2} = \frac{d}{\sqrt{b^2 + d^2}} &
\displaystyle &
\displaystyle
\sin\frac{\phi_{\sf A}}{2} = \frac{b}{\sqrt{b^2 + d^2}} &
\displaystyle &
\displaystyle
\mu_{\sf A} = -\frac{ac + bd}{b^2 + d^2}
\end{array}
\end{equation}

\begin{equation}
\label{lamda}
\lambda_{\sf A}=\frac{1}{\sqrt{c^2 + d^2}}
\end{equation}

Applying the above decomposition on the RHS of equation(2.25) we find the LHS.

The corresponding Hamiltonians to the above discrete   maps ${\sf A}\in SL(2,\mathbb{R})$  must be linear  in the generators
 $x_\mu$, of $SL(2,\mathbb{R})$~(\ref{poisson_bracket}). Moreover as they
are generators of infinitesimal Lorentz transformations $SO(1,2)$ they respect causality.

In our approach, the AdS$_2$  radial and time directions are treated as "phase space" variables, whereas time evolution is
determined by the group action. We may note that  the stringy uncertainty relations hold between the energy and the corresponding physical length or equivalently  in our case, between time and the radial extent.  As such, the interpretation of AdS$_2$ as phase space is suitable for strings moving in this background.

It is essential for the AdS/CFT holographic correspondence to define a conformal compactification of its boundary.The frequently used compactification is the conformal rescaling of the metric which gives rise to the  Poincare patch, 
covering half of the AdS$_2$ spacetime.

 Another  conformal compactification, in Minkowski signature of AdS$_2$, is obtained by stereographic projection to the $x_0=0$ plane.  Any point, $(\xi_0,\xi_1,\xi_2)\in\mathrm{AdS}_2$,
  is projected through the base point $\bm{p}$ to a point on the
 plane $x_0=0$  with coordinates $(x_1,x_2)$:
\begin{equation}
\label{stereogproj}
\begin{array}{l}
\displaystyle
x_1 = \frac{\xi_1}{1-\xi_0}\\
\displaystyle
x_2 = \frac{\xi_2}{1-\xi_0}
\end{array}
\end{equation}
Introducing the light cone coordinates of the projection plane, $x_\pm\equiv x_1\pm x_2$,  we can
parametrize AdS$_2$ as follows
\begin{equation}
\label{AdS2parametrize}
\begin{array}{l}
\displaystyle
\xi_1 = \frac{x_+ + x_-}{1+x_+ x_-}\\
\displaystyle
\xi_2 = \frac{x_+ - x_-}{1+x_+ x_-}\\
\displaystyle
\xi_0= \frac{x_+ x_- -1}{1+x_+ x_-}\\
\end{array}
\end{equation}
We observe that the stereographic projection from the point $\bm{p}=(1,0,0)$, maps each of  the light cones,
 $\bm{l}_\pm({\bf p})=\{(1,\pm\mu,\mu)|\mu\in\mathbb{R}\}$,  to two points on  the boundaries.

In order to parametrize uniquely the points on $\bm{l}_\pm(\bm{p})$, we must use the stereographic projection from the ``antipode'', $\bm{q}=(-1,0,0)$. If we call the new coordinates on the $x_0=0$ plane, $y_\pm$, we have
\begin{equation}
\label{AdS2parametrizeY}
\begin{array}{l}
\displaystyle
\xi_1 = \frac{y_+ + y_-}{1+y_+ y_-}\\
\displaystyle
\xi_2 = \frac{y_+ - y_-}{1+y_+ y_-}\\
\displaystyle
\xi_0= \frac{1-y_+y_-}{1+y_+ y_-}\\
\end{array}
\end{equation}
This coordinate system has the same problem for the points on the light cones $\bm{l}_\pm(\bm{q})$. We easily check that the stereographic projection from $\bm{p}$ (respectively  $\bm{q}$) maps the light cone axes, $x_+=0$ or $x_-=0$ of the projective plane, $x_0=0$ to $\bm{l}_\pm(\bm{q})$ (respectively $\bm{l}_\pm(\bm{p})$). More generally,  the  curves, on AdS$_2$, defined by
$x_+=\mathrm{const}$ or $x_-=\mathrm{const}$ (correspondingly for $y_\pm$) are the  light-cone  straight lines, which generate AdS$_2$.

The transition functions  between the two coordinate system are
\begin{equation}
\label{x2y}
x_- y_+ = 1 = x_+y_-
\end{equation}
 In terms of $x_\pm$,  the induced metric takes the form:
\begin{equation}
\label{lightlikemetric}
ds^2 = 4\frac{dx_+ dx_-}{(1+x_+ x_-)^2}
\end{equation}
Similarly, for $y_\pm$.

We observe now that the induced metric is invariant under the M\"obius transformations
\begin{equation}
\label{Moebius_boundary}
\begin{array}{l}
\displaystyle
x_+\to {\sf A} x_+\equiv \frac{ax_+ + b}{cx_+ +d}\\
\displaystyle
x_-\to \left[{\sf A}^{-1}\right]^\mathrm{T} x_-\equiv\frac{dx_- -c}{-bx_- +a}
\end{array}
\end{equation}
These transformations result from the Weyl action~(\ref{Weyl_mapping}) through the use  of the stereographic light cone parametrizations of AdS$_2$~(\ref{AdS2parametrize}) and~(\ref{AdS2parametrizeY}). In contrast to the other coordinate systems the variables $(x_+,x_-)$ do not mix under the isometry group.

By definition the following identity holds, for any ${\sf A}\in SL(2,\mathbb{R})$:
\begin{equation}
\label{SL2Rident}
\left[{\sf A}^{-1}\right]^\mathrm{T} = \varepsilon{\sf A}\varepsilon^\mathrm{T}
\end{equation}
where
\begin{equation}
\label{epsilon}
\varepsilon\equiv\left(\begin{array}{cc} 0 & 1 \\ -1 & 0\end{array}\right)
\end{equation}
Therefore, eq.~(\ref{Moebius_boundary})  implies that $(x_+,x_-)$ are conjugate variables and the stereographic projection plane is promoted to a phase space. Indeed, the AdS$_2$/CFT$_1$ correspondence is based on the fact that $SL(2,\mathbb{R})$  plays three different roles: (a) as the isometry group of AdS$_2$, (b) as the symplectic group of AdS$_2$ being taken  as a phase space and (c) as the conformal, M\"obius group of the boundary CFT$_1$.

The variables, $x_\pm$,  are thus appropriate holographic variables,
 because the isometry transformation group of AdS$_2$ is reduced on them to two conjugated copies of the  
1d M\"obius conformal group.

 We come now to the parametrization of the boundary, which is disconnected and consists of two circles at $x_2\to\pm\infty$. In  the covering space the boundary is
 $\mathbb{R}\times\{1,-1\}$.

Because of their transformation properties, the variables $x_\pm$ are the most suitable to use in order  to define the
two disconnected components of the  boundary, in terms of the branches of the
hyperbola (cf. eq.~(\ref{AdS2parametrize}))
\begin{equation}
\label{hyperbola_AdS2bound}
1+x_+x_- = 0
\end{equation}
This relation allows us to write $x_+$ ($x_-$) as a M\"obius transformation of the other:
\begin{equation}
\label{FourierDuality}
x_+ = -\frac{1}{x_-}\equiv \varepsilon\cdot x_-
\end{equation}
This relation is invariant under the M\"obius transformations~(\ref{Moebius_boundary}).
Therefor the two components of the boundary are two copies of the projective line $\mathbb{RP}^1$.

We notice here that the stereographic projection maps each one of the boundary components to the two branches of the hyperbola.

The boundary can also be described as the coset space, $SL(2,\mathbb{R})/\mathfrak{B}$, where $\mathfrak{B}$ is  the Borel subgroup of dilatations and translations,
\begin{equation}
\label{borelsgRP1}
\mathfrak{B}=\left\{
\left. {\sf B}(b,\lambda)=
\left(\begin{array}{cc}
\lambda & b\\ 0 & \lambda^{-1}
\end{array}\right)
\right|
\lambda\in\mathbb{R}^\ast,b\in\mathbb{R}
\right\}
\end{equation}
which preserves the point at infinity, $(x_+=\infty,x_-=0)$.

For any ${\sf A}\in SL(2,\mathbb{R})$ we have the decomposition
\begin{equation}
\label{SL2Rdecomp}
{\sf A} = {\sf R}(\phi){\sf B}(b,\lambda)
\end{equation}
and the elements ${\sf R}(\phi)\in SO(2,\mathbb{R})$ parametrize the boundary.

It will be useful later to parametrize the bulk coset representatives, $h(\phi,\mu)$ and the boundary representatives, ${\sf R}(\phi)$ by the light cone coordinates $x_\pm$. The map is the following:
\begin{equation}
\label{bulkboundLC}
\begin{array}{l}
\displaystyle
x_+ = \frac{1}{\tan\frac{\phi}{2}}\\
\displaystyle \\
\displaystyle
x_- = \frac{1-\mu\tan\frac{\phi}{2}}{\mu+\tan\frac{\phi}{2}}
\end{array}
\end{equation}
The boundary is reached when $\mu\to\pm\infty$. Indeed, a measure of the distance from the boundary is
\begin{equation}
\label{boundary_dist}
z\equiv 1+x_+x_-=\frac{2}{\sin\phi(\mu+\tan\frac{\phi}{2})}
\end{equation}
The coset representatives of the bulk and the boundary become functions $h(x_+,x_-)$ and
${\sf R}(x_+)$ respectively. So $x_+$ parametrizes motions parallel to the boundary and $x_-$ motions towards the boundary.

In order to relate the classical action of $SL(2,\mathbb{R})$ in the bulk~(\ref{one_step_op}) with the corresponding action on the boundary defined as
\begin{equation}
\label{one_step_bd}
{\sf R}(\phi_{\tau+1}) = {\sf A}{\sf R}(\phi_\tau)
\end{equation}
we must compute the RHS through the decomposition
\begin{equation}
\label{RHScoset}
{\sf A}{\sf R}(\phi_\tau) = {\sf R}(\phi_{\tau+1}){\sf B}(b_{\tau+1},\lambda_{\tau+1})
\end{equation}
and mod out the Borel factor ${\sf B}(b_{\tau+1},\lambda_{\tau+1})$.

Closing this section we explain in the following, why we have chosen to parametrize  the bulk - boundary geometry and dynamics by the corresponding group cosets.
The basic reason is their role in a new proposal for the  holographic correspondence, which we think presents an extension of the standard AdS$_2$/CFT$_1$ one.

 By construction the ${\sf R}(\phi)$ coset representatives cannot detect the distance from the boundary, i.e. $x_-$. Only at the quantum level, where an uncertainty relation, between $x_+$ and $x_-$, exists and it reflects  the UV/IR connection, is it possible from the distribution of  $x_+$ on the boundary, to get information for the distribution of $x_-$ in the bulk.

The quantum mechanical states which maximize the flow of  quantum mechanical  information between the bulk and the boundary, given the coset structure
of their geometries, are the corresponding coherent states (wavelets).

 They form overcomplete sets of states with  classical transformation properties but powerful enough to describe quantum dynamics and geometry at the same time~\cite{coherent_states}.
 We shall present the construction of these states and their properties in the next section, after we have introduced the modular discretization of the geometry and  dynamics on AdS$_2$.

\section{Modular discretization and quantum dynamics on AdS$_2[N]$}\label{HorModN}

Recent discussions on the quantum  black hole entropy of  extremal black holes and the AdS$_2$/CFT$_1$ correspondence suggest the  identification of the black hole entropy with the logarithm  of the string ground state degeneracy~\cite{Sen} .
This is an integer, $N$, fixed by the set of the black hole's  electric and magnetic charges.

Since in the Hilbert space of the degenerate ground state,  we have at most $N$ linearly independent wave functions,  the geometry resolved by the probe is fuzzy, with resolution $1/N$.

 In order to model  the  geometry and the dynamics of black hole information processing, we should  take into account  the following constraints, which have been discussed in the literature on the black hole information paradox:

\begin{itemize}

\item  In the vicinity   of the black hole horizon, the dynamics is chaotic and strongly mixing.  Any additional   bit of information, that falls into the black hole, in a very short time,  reaches (dynamic)  equilibrium with the other microscopic degrees of freedom comprising the blackhole horizon.

Furthermore, the mixing should be holographic: any subset of horizon qubits has a coarse--grained representation of the total infalling information.

This leads to the following constraints on the geometry and the dynamics:

\item Randomness, non-locality and factorization of the space-time geometry. It implies that the
total Hilbert space factorizes into a tensor product of local, coarse--grained, Hilbert spaces~\cite{Giddings,Bousso,Banks}.

\item The dynamics should provide the fastest possible quantum information processing, saturating the scrambling time bound~\cite{Page,Preskill_Hayden,Susskind_scramble,Avery,Harlow:2013tf}.

\end{itemize}

We propose to model this random, non--local and factorizable  geometry by a  number--theoretic discretization, that preserves the corresponding   group-theoretical structure of  AdS$_2$ spacetime.
This is done by replacing AdS$_2$ by the discrete cosets, AdS$_2[N] =SL(2,\mathbb{Z}_N)/SO(1,1,\mathbb{Z}_N)$. We thereby replace the set of real numbers, $\mathbb{R}$, by the set of integers modulo $N$. We call this ``modular discretization''.
This is a finite, random, set of points in the embedding Minkowski spacetime $\mathscr{M}^{2,1}$.
In the mathematical literature, such a set of points is called a {\em finite geometry}~\cite{terras,finitegeometry}.
Introducing appropriate length scales and taking the large $N$ limit we can check that the smooth geometry of AdS$_2$ emerges.

 To accommodate the above requirements on the dynamics, we employ  discrete time maps. These are  the Arnol'd cat maps,  ${\sf A}$ in $SL(2,\mathbb{Z}_N)$. These are known to exhibit strong  mixing, ergodic properties~\cite{Arnold,Vivaldi,Berry,Ford}., non-locality and factorization in the cutoff discretization parameter, $N$~\cite{fastqmaps,afnholo}.

We restrict our construction to the case $N=p$ prime for  the technical simplicity of the presentation of our arguments. In this case, the set of integers modulo $p$ is the simplest Galois field, $\mathbb{F}_p$. The unitary, irreducible, representations of the isometry group of AdS$_2[p]$, $SL(2,\mathbb{F}_p)$, are known~\cite{Silberger}.

The restriction to $N$ prime can be removed by noticing some interesting
factorizations: If $N=N_1 N_2$, with $N_{1,2}$ coprime, then we
have~\cite{fastqmaps}
\begin{equation}
\label{factorizationSL2ZN}
SL(2,\mathbb{Z}_{N_1 N_2}) = SL(2,\mathbb{Z}_{N_1})\otimes
 SL(2,\mathbb{Z}_{N_2})
\end{equation}
and
\begin{equation}
\label{factorizationAdS2N}
\mathrm{AdS}_2[N_1 N_2] = \mathrm{AdS}_2[N_1] \otimes \mathrm{AdS}_2[N_2]
\end{equation}
These factorizations imply that all powers of primes, $2^{n_1},
3^{n_2},5^{n_3},\ldots$, are the building blocks of our construction. The physical interpretation of this factorization is that the most coarse--grained Hilbert spaces on the horizon have dimensions powers of primes.

We observe that by taking tensor products over all powers of a fixed prime, $p$, we can model dynamics over the $p-$adic spacetime,  AdS$_2[\mathbb{Q}_p]$.

In order to study the finite geometry of AdS$_2[p]$, we recall the following facts about its ``isometry group''  $SL(2,\mathbb{F}_p)$:

The order   of $SL(2,\mathbb{F}_p)$ is $p(p^2-1)$. For the subgroups of rotations, ${\sf R}$, translations,
${\sf T}_\pm$ and dilatations, ${\sf D}$, the orders are $p+1,p$ and $p-1$ respectively. So the finite geometry of AdS$_2[p]$ has $p(p+1)$ points.

The set of points of the finite geometry of AdS$_2[p]$ is, by definition, the set of all solutions of the equation
\begin{equation}
\label{AdS2_p}
x_0^2 + x_1^2 -x_2^2\equiv\,1\,\mathrm{mod}\,p
\end{equation}
This can be parametrized as follows:
\begin{equation}
\label{LCAdS2N}
\begin{array}{l}
\displaystyle
x_0\equiv (a-b\,\mu)\,\mathrm{mod}\,p\\
\displaystyle
x_1\equiv  (b+a\,\mu)\,\mathrm{mod}\,p\\
\displaystyle
x_2\equiv\,\mu\,\mathrm{mod}\,p
\end{array}
\end{equation}
where $a^2 + b^2\equiv\,1\,\mathrm{mod}\,p$ and
$a, b, \mu\in\mathbb{F}_p$.

The points of AdS$_2[p]$ comprise the bulk--we must add the points on the boundary.

The boundary is the ``mod $p$'' projective line, $\mathbb{RP}_p^1$, defined as the set
\begin{equation}
\label{RP1p}
\mathbb{RP}_p^1 = GF^\ast[p]\cup\{0,\infty\}
\end{equation}
so the number of boundary points (cosets) is $p+1$.

We shall now define  the quantum mechanics of the probes  of  the bulk AdS$_2[p]$ and its boundary,
as well as the corresponding coherent states~\cite{coherent_states}.

We start with the construction of finite quantum mechanics (FQM) in the bulk. It is obvious that the set of the states and the set of observables should carry a representation of the coset structure of the bulk. We choose the  space of states to be the Hilbert space, of dimension $p$, of the metaplectic representation of $SL(2,\mathbb{F}_p)$~\cite{athanasiu_floratos}.  This choice is motivated by the fact that the spatial part of AdS$_2[p]$ is the finite field $\mathbb{F}_p$, the set of values of the space--like variable $x_-$. The wavefunctions will be the normalized  elements of the complex, projective,  space $\mathbb{CP}^{p-1}$.

 In the papers~\cite{floratos89,athanasiu_floratos,afnholo} the explicit construction of the metaplectic representation of $SL(2,\mathbb{F}_p)$ has been presented, as well as various sets of coherent states.

 The building blocks of the observables of FQM are two $p\times p$,
unitary, matrices, ``clock'', $Q$ and ``shift'', $P$,
representing the ``exponentials'' of the position and
momentum operators (for periodic boundary conditions)~\cite{Schwinger}:
\begin{equation}
\label{PandQ}
\begin{array}{ccc}
\displaystyle Q_{k,l} = \omega^k\delta_{k,l}, &
\displaystyle &
\displaystyle P_{k,l} = \delta_{k-1,l},
\end{array}
\end{equation}
 $k,l\in\mathbb{F}_p$ and $\omega=\exp(2\pi\mathrm{i}/p)$ is the $p$th root of
unity.

These matrices satisfy the exponentiated Heisenberg--Weyl commutation
relation
\begin{equation}
\label{HWcomm}
Q P = \omega P Q
\end{equation}
A useful basis is provided  by the magnetic translations
\begin{equation}
\label{HWgroup}
J_{r,s} = \omega^{r\cdot s/2} P^r Q^s,
\end{equation}
elements of the (finite) Heisenberg--Weyl group,
where the $1/2$ in the exponent is computed mod\,$p$.

The $J_{r,s}$ realize a projective representation of the translation group on the discrete torus, $\mathbb{T}_p=\mathbb{F}_p\times\mathbb{F}_p$:
\begin{equation}
\label{HWrep}
J_{r,s} J_{r',s'} = \omega^{(r's-rs)/2}J_{r+r',s+s'}
\end{equation}
they are unitary
\begin{equation}
\label{unitaryJrs}
\left[J_{r,s}\right]^\dagger = J_{-r,-s}
\end{equation}
and periodic
\begin{equation}
\label{periodicityJrs}
\left[J_{r,s}\right]^p = I_{p\times p}
\end{equation}
The phase factor in eq.~(\ref{HWrep}) is a cocycle and represents the non-commutativity of the quantized torus,$\mathbb{T}_\theta$ ($\theta=2\pi/p$)~\cite{Manin_Marcolli,Connes}.

The exact quantization of Arnol'd cat maps, ${\sf A}\in SL(2,\mathbb{F}_p)$,
is given by unitary matrices, $U({\sf A})$, satisfying
\begin{equation}
\label{metaplectic}
U({\sf A})J_{r,s}U({\sf A})^\dagger = J_{(r,s){\sf A}^{-1}}
\end{equation}
This is the definition of the metaplectic representation of
$SL(2,\mathbb{F}_p)$, which, in general, is projective.

 We can find a proper representation of
$SL(2,\mathbb{F}_p)$ which, then satisfies
the relation
\begin{equation}
\label{exactirrep}
U({\sf A})U({\sf B}) = U({\sf AB})
\end{equation}
for all ${\sf A},{\sf B}\in SL(2,\mathbb{F}_p)$. This can be done because of the following theorem: Every projective representation of $SL(2,\mathbb{F}_p)$ can be lifted to a proper representation~\cite{Weil}.

The proper representation that corresponds to the metaplectic one is given
by the following expression~\cite{balian_itzykson,athanasiu_floratos}
\begin{equation}
\label{exactirrepA}
\left[U({\sf A})\right]_{k,l} =
\frac{1}{\sqrt{p}}(-2c|p)\left\{\begin{array}{c}
1\\ -\mathrm{i}\end{array}\right\}\omega^{-\frac{ak^2 -2kl+dl^2}{2c}}
\end{equation}
for $c\not\equiv\,0\,\mathrm{mod}\,p$ and
the Jacobi symbol, $(-2c|p)=\pm 1$, depending on whether $-2c$
is a quadratic residue mod $p$ or not and the upper term between the brackets
pertains if $p=4k+1$, while the lower if $p=4k-1$.

In the case $c\equiv\,0\,\mathrm{mod}\,p$ and $a\in\mathbb{F}_p^\ast$, then
\begin{equation}
\label{c=0_A}
{\sf A} = \left(\begin{array}{cc} a & b \\ 0 & a^{-1}\end{array}\right)
\end{equation}
and
\begin{equation}
\label{c=0_U}
U({\sf A})_{k,l} =
\frac{\left(-a|p\right)}{\sqrt{p}}\left\{\begin{array}{c} 1 \\ -1\end{array}\right\}
\omega_p^{-\frac{a b}{2}k^2}\delta_{k,a^{-1}l}
\end{equation}
An important application of eq.~(\ref{exactirrepA}) is for the  Quantum Fourier Transform (QFT). For
\begin{equation}
\label{FourierS}
{\sf F}=\left(\begin{array}{cc} 0 & -1 \\ 1 & 0\end{array}\right)
\end{equation}
the corresponding unitary operator is given by
\begin{equation}
\label{FourierU}
U({\sf F}) = \frac{1}{\sqrt{p}}(-2|p)\left\{\begin{array}{c} 1 \\ -\mathrm{i}\end{array}\right\}\omega_p^{kl}=
(-2|p)\left\{\begin{array}{c} 1 \\ -\mathrm{i}\end{array}\right\}F
\end{equation}
and
\begin{equation}
\label{FourierFT}
F_{k,l}=\frac{1}{\sqrt{p}}\omega_p^{kl}
\end{equation}
is the QFT matrix.

The representation given in eq.~(\ref{exactirrepA}) is reducible: It is decomposed into two, irreducible, components~\cite{balian_itzykson}:
\begin{equation}
\label{irrepAsibspaces}
U({\sf A})_\mathrm{L,R} = U({\sf A})\frac{I_{p\times p}\pm S}{2}
\end{equation}
where
\begin{equation}
\label{projS}
S=F^2
\end{equation}
From eqs.~(\ref{FourierS}) and~(\ref{FourierU}) we deduce that
\begin{equation}
\label{Fourier4}
U({\sf F}^4) = F^4 =I_{p\times p}
\end{equation}
and, thus, the eigenvalues of $S$ are $\pm 1$, which label the chiralities of the two irreducible components. The dimension of the corresponding eigenspaces is $(p\pm 1)/2$.

It is possible to generalize the metaplectic representation from the  discretization  $N=p$ prime to any integer $N$ by noting that, if $N$ is composite, $N=N_1 N_2$, with $N_1,N_2$ coprimes, for
every ${\sf A}\in SL(2,\mathbb{Z}_{N_1N_2})$ we obtain
\begin{equation}
\label{Afactorization}
{\sf A} = {\sf A}_1\cdot{\sf A}_2
\end{equation}
with ${\sf A}_i\in SL(2,\mathbb{Z}_{N_i})$, $i=1,2$.

It can be proved that the unitary matrix $U({\sf
  A})$ of eq.~(\ref{exactirrepA}) ``inherits'' this property as follows
\begin{equation}
\label{Ufactorization}
U({\sf A}) = U({\sf A}_1)\otimes U({\sf A}_2)
\end{equation}
The $N_1N_2\times N_1N_2$ matrix $U({\sf A})$ decomposes into a tensor product
of an $N_1\times N_1$ and an $N_2\times N_2$ unitary matrix. This leads to an
acceleration of the computation of the action of the quantum map $U({\sf A})$
on the Hilbert space of states, ${\cal H}_{N_1N_2}$, from $O(N^2)$ to $O(N\ln
N)$ operations~\cite{fastqmaps}.

Thus,  the building blocks of FQM are the Hilbert spaces  of dimension
$N=p^n$, with $p$ an odd prime and $n\in\mathbb{N}$.

\section{Bulk and boundary coherent states}\label{CohStates}
 
The coherent state method selects an invariant state under the stability group, as the ground state, $|0\rangle_{\sf D}$~\cite{Grosse}.

 For the bulk the stability group is the scaling group, ${\sf D}$.
 The corresponding quantum map, $U({\sf D}(\lambda))$, for $\lambda\in\mathbb{F}_p^\ast$, is the circulant matrix:
\begin{equation}
\label{circulant_matrix}
U({\sf D}(\lambda))_{k,l} = \left(-\lambda|p\right)\left\{\begin{array}{c} 1 \\ -1\end{array}\right\}
\delta_{k,\lambda^{-1}l}
\end{equation}

We choose as ground state a common eigenvector of $U({\sf D}(\lambda))$ for all $\lambda$, namely
\begin{equation}
\label{groundstate0D}
|0\rangle_{\sf D} = \frac{1}{\sqrt{p}}\left(\underbrace{1,1,\ldots,1}_{p}\right)
\end{equation}

The coherent states for AdS$_2[p]$ are now defined as
\begin{equation}
\label{coherent_states}
|h\rangle = U({\sf h})|0\rangle_{\sf D}
\end{equation}
for all $h\in\mathrm{AdS}_2[p]$
and $U(h)$ the $p\times p$ unitary matrix constructed in eq.~(\ref{c=0_U}).

We notice here that the vacuum $|0\rangle_{\sf D}$ is annihilated by the projector $P_-=(I-S)/2$. In effect it belongs to the subspace of the projector, $P_+=(I+S)/2$, which has dimension $p_+\equiv (p+1)/2$.
The matrix $S$ commutes with all matrices $U({\sf h})$. It implies that the coherent states $|h\rangle$  belong 
to the eigenspace of $P_+$. This is the positive chirality eigenspace. It is possible to construct coherent states, belonging to the orthogonal eigenspace  of dimension $p_-=(p-1)/2$, by choosing the common eigenstate of the dilatation group among the eigenvectors of $S$ with opposite chirality.

We can use the parametrization of the cosets by rotations and translations in order to obtain explicit expressions for the coherent states, $|h\rangle$:

For
\begin{equation}
\label{hcoset}
h(\phi,\mu)=\left(\begin{array}{cc} a & -b \\ b & a\end{array}\right)
                   \left(\begin{array}{cc} 1 & -\mu\\ 0 & 1\end{array}\right)
\end{equation}
with $a^2+b^2\equiv 1\,\mathrm{mod}\,p$. Using eqs.~(\ref{bulkboundLC}) we find the relation between
$a,b,\mu$ and $x_\pm$, namely
\begin{equation}
\label{bulkboundLCxpm}
\begin{array}{c}
\displaystyle
x_+ = \frac{a}{b}\\
\displaystyle
x_- = \frac{a-b\mu}{a\mu+b}
\end{array}
\end{equation}
In components:
\begin{equation}
\label{coherent_states_h}
\left\langle k|h\right\rangle =
\frac{1}{\sqrt{p}}\left((a-b\mu))|p\right)
\omega_p^{\frac{b+\mu a }{2(a-b\mu)} k^2} =
\frac{1}{\sqrt{p}}\left((a-b\mu))|p\right)
\omega_p^{\frac{k^2}{2x_-}}
\end{equation}
where $k=0,1,2,\ldots,p-1$.
The coherent states $|h\rangle$ of the bulk, can be, therefore, parametrized in terms of $x_\pm$ which will be denoted by $|x_+,x_-\rangle$.

These definitions imply the classical transformation of coherent states under the isometry group, namely
\begin{equation}
\label{classicaltransfcohstates}
U({\sf A})|h\rangle = |{\sf A}h\rangle
\end{equation}
These states form an overcomplete set of normalized states with very useful properties:
\begin{itemize}
\item
 Resolution of the identity:
\begin{equation}
\label{resol_ident}
\frac{1}{d}\sum_{h}|h\rangle\langle h| =  P_+
\end{equation}
where $d$ is defined
\begin{equation}
\label{resol_ident_state}
d = \sum_{h} \left| \left\langle h| h_1\right\rangle\right|^2
\end{equation}
for any state $|h_1\rangle$.

The above identity is based on the irreducibility of the metaplectic representation on the subspace of positive chirality.

\item Propagator property for the function
\begin{equation}
\label{propagator}
\Delta(h_1,h_2) = \left\langle h_2|h_1\right\rangle
\end{equation}
This function has the property of a ``reproducing kernel'' (propagator)
\begin{equation}
\label{reprokernel}
\Delta(h_1,h_2)=\frac{1}{d}\sum_h\Delta(h_1,h)\Delta(h,h_2)
\end{equation}
and is invariant under the isometry group.

\item For a general state $|\psi\rangle$ we have
\begin{equation}
\label{genstatedecomp}
|\psi\rangle = \frac{1}{d}\sum_h |h\rangle\langle h|\psi\rangle
\end{equation}
\item The symbol of operators: To an operator $\widehat{A}$ we associate a scalar function
$\widetilde{A}(h_1,h_2)$ with
\begin{equation}
\label{opsymbol}
\widetilde{A}(h_1,h_2)\equiv \langle h_2|\widehat{A}|h_1\rangle
\end{equation}
so that
\begin{equation}
\label{opsymbolinv}
\widehat{A}=\frac{1}{d^2}\sum_{h_1,h_2}\widetilde{A}(h_1,h_2)|h_2\rangle\langle h_1|
\end{equation}
For two operators $\widehat{A},\widehat{B}$ we assign as symbol of their product the expression
\begin{equation}
\label{symbolprod}
\widetilde{AB}(h_1,h_2)=\frac{1}{d}\sum_h \widetilde{A}(h_1,h)\widetilde{B}(h,h_2)
\end{equation}
\end{itemize}
The local quantum observables, that have   ''nice'' transformation properties, are the $p\times p$ hermitian matrices $Q_I(h)$, such that
\begin{equation}
\label{Qvectobs}
U(\textsf{A})Q_I(h) U(\textsf{A})^\dagger=R_{IJ}Q_J(\textsf{A}h),\, I,J=1,\ldots,\mathrm{dim}\,R
\end{equation}
For scalar observables (scalar fields) we must have
\begin{equation}
\label{Qscalobs}
U(\textsf{A})Q(h) U(\textsf{A})^\dagger= Q(\textsf{A}h)
\end{equation}
The simplest ones are the (pure state) density matrices,
\begin{equation}
\label{density_matrix}
\rho(h) = |h\rangle\langle h|
\end{equation}
that can be used as a basis for measurement.

For any scalar function, $f(h)$, on AdS$_2[p]$, we can construct the quantum observable
\begin{equation}
\label{Qobs}
\mathcal{O}(f)\equiv \sum_{h}f(h)|h\rangle\langle h|
\end{equation}
which is hermitian if $f(\cdot)$ is real.

The one--time--step evolution of these observables is given by
\begin{equation}
\label{1tstepevol}
\mathcal{O}_{n+1}(f) = U(\textsf{A})\mathcal{O}_n(f)U(\textsf{A})^\dagger
\end{equation}
with initial condition $\mathcal{O}_{n=0}(f)=\mathcal{O}(f)$.

We may write this relation in the following way:
\begin{equation}
\label{1tstepevol1 }
\mathcal{O}_{n+1}(f) = \mathcal{O}_n\left(f\circ \textsf{A}^{-1}\right)
\end{equation}
The set of time correlation functions for these observables defines FQM on AdS$_2[p]$:
\begin{equation}
\label{tcorrfuns}
G(t_1,t_2,\ldots,t_n|f_1,f_2,\ldots,f_n) =
_{\sf D }\hskip-0.16truecm\left\langle 0|
\mathcal{O}_{t_1}(f_1)\mathcal{O}_{t_2}(f_2)\ldots \mathcal{O}_{t_n}(f_n)|0\right\rangle_{\sf D}
\end{equation}
We shall present the bulk/boundary correspondence for the quantum  map dynamics using the parametrization of both spaces by the light cone variables, $x_\pm$, of the stereographic projection. The coherent state $\left |h\right\rangle\equiv\left|x_+,x_-\right\rangle$ and the action of an $SL(2,\mathbb{F}_p)$ group element $\textsf{A}$ will be lifted to the action of the unitary operator, $U({\sf A})$ as follows:
\begin{equation}
\label{ group_actionbB}
U({\sf A})\left|x_+,x_-\right\rangle = \left|\frac{a x_+ +b}{c x_+ +d},\frac{dx_- -c}{a-bx_-}\right\rangle
\end{equation}

Let us now pass to the construction of the coherent states and the observables on the boundary.

The boundary ${\sf Bd}[p]$  of the discrete space-time  $\mathrm{AdS}_2[p]$  is defined, in analogy with the continuum case, by conformal compactification. In light--cone coordinates, $(x_+,x_-)$, of the projective plane, it is described as the set of points
\begin{equation}
\label{BdN}
1+x_+x_-\equiv 0\,\mathrm{mod}\,p
\end{equation}
For every $x_+\in\mathbb{F}_p^\ast$, we have $x_-\equiv -x_+^{-1}\,\mathrm{mod}\,p$. We must also add the points at ``infinity'', $x_+=0,x_-=\infty$ and $x_+=\infty,x_-=0$. So the boundary comprises the $p+1$  points
\begin{equation}
\label{BdNinfty}
{\sf Bd}[p] =\left\{
\left.\left(x_+,-x_+^{-1}\right)\right|
x_+\in\mathbb{F}_p^\ast
\right\}
\cup
\left\{(\infty,0),(0,\infty)\right\}
\end{equation}
The boundary set is invariant under the M\"obius group, $SL(2,\mathbb{F}_p)$. It is the  coset space $SL(2,\mathbb{F}_p)/\mathfrak{B}_p$, where $\mathfrak{B}_p$ is the Borel subgroup  that preserves the ``point at infinity'', $q=(x_+=\infty,x_-=0)$:
\begin{equation}
\label{borelsg}
\mathfrak{B}_p=\left\{
\left.
\left(\begin{array}{cc}
\lambda & b\\ 0 & \lambda^{-1}
\end{array}\right)
\right|
\lambda\in\mathbb{F}_p^\ast,b\in\mathbb{F}_p
\right\}
\end{equation}
In the $p-$dimensional Hilbert space of the metaplectic representation the quantum maps, corresponding to $\mathfrak{B}_p$,  are given as
\begin{equation}
\label{Uborelsg}
U\left[
\left(\begin{array}{cc}
\lambda & b\\ 0 & \lambda^{-1}
\end{array}\right)
\right]_{k,l}=
\left(-\lambda|p\right)\left\{\begin{array}{c} 1 \\ -1\end{array}\right\}
\omega_p^{-\frac{\lambda b}{2}k^2}\delta_{k,\lambda^{-1}l}
\end{equation}
The ``vacuum''  on which we define  the  coherent states of the boundary must be a common eigenvector of the stability group $\mathfrak{B}_p$~\cite{coherent_states}. We can check from~(\ref{Uborelsg}) that there is only one such eigenvector, $|0\rangle_q$:
\begin{equation}
\label{vacuum}
\left\langle k| 0\right\rangle_q = \delta_{k,0}
\end{equation}
The subgroup of $SL(2,\mathbb{F}_p)$, that acts transitively on the boundary is generated by the rotation subgroup $SO(2,\mathbb{F}_p)$:
\begin{equation}
\label{boundary_transit}
SO(2,\mathbb{F}_p)=\left\{
\left.
\left(\begin{array}{cc} a & -b \\ b & a\\ \end{array}\right)
\right|
a^2+b^2\equiv 1\,\mathrm{mod}\,p
\right\}
\end{equation}
This is an abelian, cyclic,  subgroup of order $p+1$, if $p=4k-1$ and $p-1$ if $p=4k+1$. The generator of this cyclic group can be found by random search~\cite{athanasiu_floratos}.

The discrete coherent states of the boundary ${\sf Bd}[p]$ can now be defined as
\begin{equation}
\label{discrete_coherent_states}
| x_+\rangle = U({\sf R}(x_+))|0\rangle_q
\end{equation}
From~(\ref{exactirrepA}) we obtain the expression for the ground state wave function, $\langle k|x_+\rangle$ as
\begin{equation}
\label{groundstate}
\langle k|x_+\rangle = \frac{1}{\sqrt{p}}(-2b|p)\left\{\begin{array}{c} 1\\ \mathrm{i}\end{array}\right\}
\omega_p^{-\frac{x_+k^2}{2}}
\end{equation}
when $x_+\in\mathbb{F}_p^\ast$. For the two additional points of the boundary, $x_+=\infty,x_-=0$(then $b=0$) and $x_+=0,x_-=\infty$ (then $a=0$) the corresponding coherent states are:

\begin{itemize}
\item
When $b=0$, we need, in principle,  to distinguish two cases: $a=+1$ and $a=-1$--but,  since the action of the group is projective, these lead to the same state,
 $|x_+=\infty\rangle = |0\rangle_q$, whence $\langle k|x_+\rangle = \delta_{k,0}$.
\item
For $a=0$, then $b=-1$, $x_+=0$ The state $|x_+=0\rangle$ leads to the constant wavefunction
\begin{equation}
\label{xplusinfty}
\langle k| x_+\rangle = \frac{1}{\sqrt{p}}\left(2|p\right)\left\{\begin{array}{c} 1\\ \mathrm{i}\end{array}\right\}
\end{equation}
for all $k=0,1,\ldots,p-1$.
\end{itemize}
In total we get $p+1$ states, matching the number of points on the boundary.

In analogy with the bulk coherent states we observe that the ground state $|0\rangle_q$  of the boundary is annihilated by the projector $(I-S)/2$, as are all the coherent states $|x_+\rangle$.
Thus, these coherent states live in the eigenspace of $P_+=(I+S)/2$, which is $(p+1)/2-$dimensional. They form an overcomplete set of states and display all the expected features of coherent states.

Let us now turn our attention to the boundary observables.  We construct the following class of operators, which have nice transformation properties under the M\"obius conformal group and form a basis for measurement.

Using the magnetic translations, $J_{r,s}$, of the Heisenberg--Weyl group, where $r,s=0,1,2,\ldots,p-1$, we define the operators
\begin{equation}
\label{Oxplus}
\mathcal{O}(x_+)=\frac{1}{p}\sum_{s=0}^{p-1} J_{s(1,-x_+)}
\end{equation}
with $x_+=0,1,\ldots,p-1$. Their matrix elements are
\begin{equation}
\label{Oxplusmatrixels}
\left[\mathcal{O}\right]_{k,l}=\frac{1}{p}\omega_p^{-\frac{x_+(k^2-l^2)}{2}}
\end{equation}
These operators are projectors:
\begin{equation}
\label{Oxplusproj}
\begin{array}{l}
\displaystyle
\mathcal{O}(x_+)^2 = \mathcal{O}(x_+)\\
\displaystyle
\mathcal{O}(x_+)^\dagger = \mathcal{O}(x_+)
\end{array}
\end{equation}
and they transform conformally  (for all ${\sf A}\in SL(2,\mathbb{Z}_N)$):
\begin{equation}
\label{conformalOxplus}
U({\sf A})\mathcal{O}(x_+)U({\sf A})^\dagger = \mathcal{O}\left(\frac{ax_+ +b}{cx_+ +d}\right)
\end{equation}
For example, under the Fourier transform,
$$
{\sf S}=\left(\begin{array}{cc} 0 & -1 \\ 1 & 0\\ \end{array}\right)
$$
we find that
\begin{equation}
\label{FourierT}
U({\sf S})\mathcal{O}(x_+)U({\sf S})^\dagger = \mathcal{O}\left(-\frac{1}{x_+}\right)
\end{equation}
We use eq.~(\ref{FourierT}) to {\em define}
\begin{equation}
\label{Oxplusinfty}
{\mathcal O}(x_+=\infty)\equiv U({\sf S}){\mathcal O}(0)U({\sf S})^\dagger = |0\rangle_q { }_q\langle 0|
\end{equation}
In the notation of eq.~(\ref{discrete_coherent_states}) the state $|\infty\rangle$ is the ground state, $|0\rangle_q$.

The operators $\mathcal{O}(x_+)$ have the nice property that they are projectors on the discrete coherent states,
$|x_+\rangle$.
One can, indeed, check that
\begin{equation}
\label{Oxplus1}
\mathcal{O}(x_+)=|x_+\rangle\langle x_+|
\end{equation}
This holds for all $x_+=0,1,2,\ldots,p-1,\infty$.

The boundary observables can, therefore, be expressed in the $\mathcal{O}(x_+)$ basis: 
To any function, $f:{\sf Bd}[N]\to \mathbb{C}$, we assign the observable
\begin{equation}
\label{Of}
\mathcal{O}(f)=\sum_{x_+}f(x_+)\mathcal{O}(x_+)
\end{equation}
Their transformation properties are
\begin{equation}
\label{Oftransform}
\begin{array}{l}
\displaystyle
U({\sf A}) \mathcal{O}(f)U({\sf A})^\dagger =
\mathcal{O}(f\circ{\sf A}^{-1})
\end{array}
\end{equation}
In this way we may establish contact between  modular functions and forms of  the finite M\"obius group $SL(2,\mathbb{F}_p)$, and  conformal operators of definite conformal weight~\cite{terras,AxFlNic}.

Once we have defined appropriate conformal operators, it is possible to calculate their correlation functions
 (and the identities these satisfy) in any state. The two-- and three--point functions can be determined from conformal invariance; the higher--point functions depend strongly on the quantum dynamics~\cite{dAFF,chamonetal}.

In the next section we shall reconstruct bulk observables, when the boundary observables are known~\cite{Hamilton:2006az,Papadodimas:2012aq}

\section{AdS$_2[N]$/CFT$_1[N]$  coherent state holography}\label{Hologcat}

In this section we present a new AdS$_2$/CFT$_1$ correspondence, AdS$_2[p]$/CFT$_1[p]$,
based on the coherent states of positive chirality, in the bulk and the boundary. A similar method can be applied to the subspace of negative chirality.

By CFT$_1[p]$ we understand the quantum mechanics on the discrete, projective line, $\mathbb{RP}_p^1$, defined by the evolution operator $U({\sf A})$, for ${\sf A}\in SO(2,\mathbb{F}_p)$. In analogy to
the conformal quantum mechanics of ref.~\cite{dAFF}, the generator of this group corresponds to their ``second'' Hamiltonian, which has discrete spectrum. From the point of view of radial observers of the AdS$_2$ near horizon geometry of an extremal black hole, this evolution corresponds to that of freely infalling observers~\cite{Strominger,townsendetal}.

To motivate the use of coherent states for the correspondence we notice the following:

The basic strength of the AdS/CFT correspondence relies on two important facts: First, the conformal boundary completion of the AdS space-time is very specific in selecting those boundary observables, which are appropriate for the reconstruction of those in the bulk--and this is holography. The second one is more constraining in that the AdS/CFT holography satisfies a new uncertainty principle, the IR/UV connection, which is a stringy effect. The higher the energy of the probe of a system on the boundary, in order to localize it, the bigger the distance form the boundary of the gravity dual system in the bulk. In the language of the stringy uncertainty principle, the higher the energy of the closed string state in the bulk, the larger the average length of the string: $\Delta x_+ \Delta x_-\geq 1/\alpha'$.

In the light cone coordinates, $(x_+, x_-)$, on AdS$_2$, $x_+$ is parallel to the boundary and $x_-$ is a measure of the distance from it. Strictly speaking, the appropriate  quantity is $z\equiv 1+x_+x_-$, so, for fixed $x_+$, $x_-\to -1/x_+$, when $z\to 0$.

In section~\ref{LCWeyl} we observed that the variables $(x_+,x_-)$ are appropriate holographic coordinates for the bulk, since they transform under the isometry group by M\"obius, conformal, transformations:
\begin{equation}
\label{Moebius_conformal}
\begin{array}{l}
\displaystyle
x_+\to {\sf A} x_+\equiv \frac{ax_+ + b}{cx_+ +d}\\
\displaystyle
x_-\to \left[{\sf A}^{-1}\right]^\mathrm{T} x_-\equiv\frac{dx_- -c}{-bx_- +a}
\end{array}
\end{equation}
Notice that $\left[{\sf A}^{-1}\right]^\mathrm{T}$ is the Fourier transform of ${\sf A}\in SL(2,\mathbb{Z}_N)$,
\begin{equation}
\label{Fourier_group_transform}
\left[{\sf A}^{-1}\right]^\mathrm{T} = \varepsilon{\sf A}\varepsilon^\mathrm{T}
\end{equation}
So $(x_+,x_-)$ are conjugate variables, similar to position and momentum. Indeed, for AdS$_2$, they represent time and length, promoting AdS$_2$ into a stringy phase space. To saturate the stringy uncertainty principle we must employ the corresponding coherent states. In the bulk they have been defined as $|x_+,x_-\rangle$. The coordinates denote the center of the coherent state on the boundary as $|x_+\rangle$.

The bulk coherent states are the discrete analogs of the well--known wavelets, used in signal processing~\cite{Mallat}, which determine the spectrum of scales of a given signal as a function of position.

The boundary coherent states are, also,  the discrete analogs of the usual coherent states of the harmonic oscillator (albeit on a different vacuum state)~\cite{balian_itzykson,athanasiu_floratos}.

We shall describe now the reconstruction method of the bulk observables (states) from appropriate boundary ones, using the wavelet representation and its properties.

Let us choose, for any value of the variable $x_+$, an independent variable, $x_-$, which takes values on the projective line, $\mathbb{RP}_N^1$, and define the state
\begin{equation}
\label{xtildeminus}
|\widetilde{x}_-\rangle\equiv F|x_-\rangle
\end{equation}
with $ F$ the finite Fourier transform~(\ref{FourierFT}). Since $|x_-\rangle$ is a boundary coherent state, we deduce that
\begin{equation}
\label{conformalFourier}
\widetilde{x}_-=-\frac{1}{x_-}
\end{equation}
In section~\ref{CohStates} we constructed the chiral scalar operators ${\mathcal O}(x_+)$. It is obvious that the scalar operator,
\begin{equation}
\label{Oxminus}
\widetilde{{\mathcal O}}(x_-)\equiv{\mathcal O}(\widetilde{x}_-)
\end{equation}
has conjugated transformation properties, i.e.
\begin{equation}
\label{Oxminuschirality}
U({\sf A})\widetilde{{\mathcal O}}(x_-)U({\sf A})^\dagger = \widetilde{\mathcal O}\left(
\left[({\sf A}^{-1}\right]^{\mathrm{T}}x_-
\right)
\end{equation}
for any ${\sf A}\in SL(2,\mathbb{F}_p)$.

We observe now that the composite operator,
\begin{equation}
\label{compositeO}
{\mathcal O}(x_+,x_-)\equiv {\mathcal O}(x_+)\widetilde {\mathcal O}(x_-)
\end{equation}
 is a scalar operator in the bulk. Indeed,
\begin{equation}
\label{bulk_scalar}
U({\sf A}){\mathcal O}(x_+,x_-)U({\sf A})^\dagger =
{\mathcal O}({\sf A}x_+,\left[{\sf A}^{-1}\right]^{\mathrm{T}}x_-)
\end{equation}
We shall use these operators to reconstruct Hermitian bulk scalar operators, so we must symmetrize the 
product in~(\ref{compositeO}).

On the boundary, the operators ${\mathcal O}(x_+,x_-=-1/x_+)={\mathcal O}(x_+)$.

The reconstruction of the bulk operators from boundary data will be described next.
The bulk/boundary correspondence in our construction is based on the fact that the Hilbert space of the bulk coincides with the Hilbert space of the boundary and carries the positive chirality component of the metaplectic
representation of $SL(2,\mathbb{F}_p)$. This places constraints on the algebra of observables on both sides of the correspondence.

Since both the bulk and boundary coherent states are overcomplete systems in the eigenspace of $P_+$, , we get the relation
\begin{equation}
\label{bulkboundcohstat}
|x_+,x_-\rangle = \frac{1}{d}\sum_{y_+}K(x_+,x_-|y_+)|y_+\rangle
\end{equation}
where the bulk/boundary propagator, $K(x_+,x_-|y_+)$, can be explicitly calculated:
\begin{equation}
\label{bulkboundprop}
K(x_+,x_-|y_+) = \langle y_+|x_+,x_-\rangle
\end{equation}
From eqs.~(\ref{coherent_states_h}) and~(\ref{groundstate}) we find that
\begin{equation}
\label{Kxplusxminusyplus}
K(x_+,x_-|y_+) = \left((a-b\mu)|p\right)(-2b'|p)\left(\left.\frac{1}{2}\left(y_++\frac{1}{x_-}\right)\right|p\right)
\end{equation}
In this expression
\begin{equation}
\label{bulk_data}
\begin{array}{l}
\displaystyle
a=\frac{x_+}{\sqrt{x_+^2+1}} \\
\displaystyle
b = \frac{1}{\sqrt{x_+^2+1}}\\
\displaystyle
\mu=\frac{x_+-x_-}{1+x_+x_-}
\end{array}
\end{equation}
for  the bulk coherent states and
\begin{equation}
\label{boundary_data}
\begin{array}{l}
\displaystyle
a' = \frac{2y_+}{1+y_+^2}\\
\displaystyle
b' = \frac{1-y_+^2}{1+y_+^2}
\end{array}
\end{equation}
for the boundary ones.

The normalization constant, $d$, is defined through the overcompleteness relation of the boundary coherent states
\begin{equation}
\label{overcompletebound}
\sum_{y_+}|y_+\rangle\langle y_+| = d P_+
\end{equation}
Using eq.~(\ref{groundstate}) we find that
\begin{equation}
\label{cohstatenorm}
d\langle l |P_+| m\rangle = \sum_{y_+}\langle l | y_+\rangle \langle y_+ | m\rangle \Rightarrow d = 2
\end{equation}
The range of the bulk variables, $x_\pm$, is determined by the light cone parametrization of the bulk, while the range of the boundary variable, $y_+$ runs over the projective line, $\mathbb{RP}_N^1$, i.e.
$y_+\in\{0,1,2,\ldots,p-1\}\cup\{\infty\}$.

The correspondence  between bulk/boundary observables can be constructed through the relation
\begin{equation}
\label{bulkboundobs}
|x_+,x_-\rangle\langle x_+,x_-| = \frac{1}{d^2}\sum_{y_+,y_-}G(x_+,x_-|y_+,y_-){\mathcal O}(y_+,y_-)
\end{equation}
The coefficient function, $G(x_+,x_-|y_+,y_-)$, can be determined from the bulk/boundary propagator
\begin{equation}
\label{bulkboundpropG}
G(x_+,x_-|y_+,y_-) = \frac{K(x_+,x_-|y_+)K^\ast(x_+,x_-|-1/y_-)}{\langle y_+|-1/y_-\rangle}
\end{equation}
The denominator is, in fact, a boundary/boundary propagator, whereas the numerator is the product of bulk/boundary, resp. boundary/bulk propagators.

\section{Summary and conclusions }\label{Concl}

Before we summarize our results, let us review  the conditions satisfied  by our finite, discrete, model of the black hole near horizon geometry for the radial and temporal directions:
\begin{itemize}
\item 
	We managed to single out the proposed type of modular discretization in the space of all possible finite geometries, by imposing the additional condition of existence of a  holographic correspondence. This is to be satisfied through the replacement of AdS$_{2}$ by AdS$_2[N]$ and its boundary by CFT$_1[N]=\mathbb{RP}_N^1$,  . 
\item
Indeed,  the finite geometry inherits the symmetry properties of its continuous counterpart (isometry group, coset structure and its quantum representations as well as bulk-boundary correspondence) albeit in a discretized disguise.
\item
In the  framework of the specific finite geometry it is very natural to choose as a model for the dynamics of probes
the isometry group elements which, interestingly, possess strongly chaotic mixing properties.  They are the well 
known Arnol'd cat maps,  defined as M\"obius transformations on the stereographic lightcone plane.
\item
Moreover, special properties of the modular representations guarantee the factorization with respect to the
ultraviolet cut-off $N$. This is important for fast quantum information processing on the near horizon region. As we
plan to show in forthcoming works the proposed framework is capable of providing an example for  a mechanism for the saturation of the fast scrambling conjecture.\cite{susskindscramble,Barbon,Preskill_Hayden}.
\end{itemize}

In the present  work we have studied the modular discretization of the AdS$_2$ geometry, AdS$_2[N]$, and the ensuing classical and quantum dynamics of probes  using generalized Arnold cat maps. 
We have demonstrated that our toy model is successful in realizing all of the properties  which are considered key ingredients for a departure from semiclassical and local physics, namely those of  non-locality, chaotic dynamics and fast quantum information processing.

With the discretization parameter, $N$, which provides both an ultraviolet and an  infrared cutoff,   the coset space nature  of the AdS$_2$  geometry ``carries over'' at the discretized level.  The corresponding, effective Planck constant, $\hbar=2\pi/N$ can be identified, also, with the non commutativity parameter of the quantum coset geometry.\cite{Grosse}

The strong arithmetic chaos of the Arnol'd cat map dynamics is inherited in a transparent way by the coset quantum states, which are the coherent states  of $SL(2,\mathbb{Z}_N)$. It is rather interesting that there is a correspondence between the bulk and the boundary states and observables of AdS$_2[N]$;  the latter belong to  the discrete projective line, $\mathbb{RP}_N^1$. In a unique Hilbert space of finite dimension and given chirality,  by using the overcompleteness of the corresponding coherent states of the bulk and the boundary,  we provided a method to reconstruct the bulk states and observables from the corresponding boundary data. To this end we constructed the bulk--bulk, bulk--boundary and boundary--boundary propagators, which are invariant under the isometries of AdS$_2[N]$. They are given by the overlap amplitudes between the corresponding coherent states.

These propagators realize the UV/IR connection between the bulk and the boundary scales, since the corresponding coherent states saturate the string uncertainty relation  
$\Delta x_+ \Delta x_-\geq 1/\alpha'$.

Our present work can be a basis for further extensions: 

\begin{enumerate}
 \item	 In the study of  the AdS$_2[N]$/CFT$_1[N]$ correspondence for different representations of the discrete isometry group, 
$SL(2,\mathbb{Z}_N)$~\cite{AxFlNic}. In particular, it is interesting to study the modular 
discretization of the boundary conformal quantum mechanics of ref.~\cite{dAFF, chamonetal}. It requires at the group level the definition of primary operators,  their dimensions, as well as their fusion algebra.
 \item Since the classical Arnol'd cat maps possess factorization in the parameter $N$ and strong chaotic properties by choosing  $ N=p^{n}$ where $p$  is a prime integer,  we can construct the corresponding p-adic dynamics at  both the classical and quantum levels. Indeed all of our amplitudes possess factorization properties. Therefore by taking their infinite product over $n$ from 1 to infinity it is possible to construct the corresponding p-adic amplitude~\cite{p-adics}. In recent works by Barb\'on {\em et al.}~\cite{Barbon}.  it has been shown that ultrametric or p-adic  structures of the Hilbert space of black hole microstates which are supported by specific expander graphs guarantee the saturation of scrambling time bound for the black hole information processing \cite{susskindscramble}. 
  \item Since the quantum Arnol'd cat maps possess factorization in the parameter $N$ and strong chaotic properties~\cite{Barbon}, they are also appropriate for the construction of  quantum circuit models of deterministic chaos for the qubit information processing in black holes~\cite{susskindbhComp,Giddings,Preskill_Hayden,Avery,Harlow:2013tf}.
In analogy with the quantum circuit realization of Shor's factorization algorithm~\cite{Shor}, it is expected that quantum circuits for the quantum Arnol'd cat maps will provide  similar (exponential) improvements over their  classical factorization properties  and may saturate the scrambling bound~\cite{AxFlNic}, as well.
\end{enumerate}

\vskip1.5truecm

{\bf Acknowledgements:} MA and SN acknowledge the warm hospitality of the CERN Theory Division as well as of the LPTENS. We are thankful to  A. Dabholkar, M.Porrati  and E. Rabinovici for discussions and their constructive comments. EF is especially grateful to L. Alvarez-Gaum\'e, S.Ferrara and I. Antoniadis for making his stay at CERN such a rewarding experience.   
The research of EF is
implemented under the ``ARISTEIA-I'' action (Code no. 1612,  D.654) and title ``Holographic Hydrodynamics''  of the ``operational programme
education and lifelong learning'' and is co-funded by the European Social Fund (ESF) and National Resources. The research of MA was
supported in part by the General Secretariat for Research and Technology of Greece and the European Regional Development Fund 
MIS-448332-ORASY(NSRF 2007-13 ACTION, KRIPIS).

\end{document}